\documentclass[preprint,12pt]{aastex}
\usepackage{emulateapj5}

\begin{document}
\shorttitle{SiO 5\too 4 survey of low mass outflows}
\shortauthors{Gibb et al.}

\newcommand{\intdv}{$\int$\trstar$\d v$}
\newcommand{\hcop}{HCO$^{+}$}
\newcommand{\kms}{$\,$km$\,$s$^{-1}$}
\newcommand{\mic}{$\mu$m}
\newcommand{\cucm}{cm$^{-3}$}
\newcommand{\sqcm}{cm$^{-2}$}
\newcommand{\degs}{$^{\circ}$}
\newcommand{\msun}{M$_{\odot}$}
\newcommand{\lsun}{L$_{\odot}$}
\newcommand{\tastar}{$T_{\rm A}^{*}$}
\newcommand{\trstar}{$T_{\rm R}^{*}$}
\newcommand{\trad}{$T_{\rm R}$}
\newcommand{\amm}{NH$_{3}$}
\newcommand{\too}{$\rightarrow$}
\newcommand{\ceo}{${\rm C}^{18}{\rm O}$}
\newcommand{\thco}{$^{13}$CO}
\newcommand{\twco}{$^{12}$CO}
\newcommand{\ci}{C\,{\sc I}}

\title{A survey of SiO 5$\rightarrow$4 emission towards outflows from
  low-luminosity protostellar candidates}

\author{Andy G. Gibb}
\affil{Department of Astronomy, University of Maryland, College Park,
  MD 20742, USA}
\email{agg@astro.umd.edu}
\author{John S. Richer}
\affil{Cavendish Astrophysics Group, Department of Physics, University
  of Cambridge, Madingley Road, Cambridge, CB3 0HE, UK}
\email{jsr@mrao.cam.ac.uk}
\author{Claire J. Chandler}
\affil{National Radio Astronomy Observatory\footnote{The National
Radio Astronomy Observatory is a facility of the National Science
Foundation operated under cooperative agreement by Associated
Universities, Inc.}, PO Box O, Socorro, NM 87801, USA}
\email{cchandle@nrao.edu}
\and
\author{Chris J. Davis}
\affil{Joint Astronomy Centre, 660 N. A'oh\=ok\=u Place, Hilo, HI
  96725, USA}
\email{c.davis@jach.hawaii.edu}

\begin{abstract}
  
  We have observed the SiO $J$=5\too 4 line towards a sample of 25
  low-luminosity (L$_* < 10^3$ L$_\odot$) protostellar outflow
  systems.  The line was detected towards 7 of the 25 sources, a
  detection rate of 28 per cent. The majority (5 out of 7) of sources
  detected were of class 0 status. We detected a higher fraction
  of class 0 sources compared with the class I and II sources,
  although given the small numbers involved the significance of this
  result should be regarded as tentative.  Most of the detected
  sources showed emission either at or close to the central position,
  coincident with the protostar. In four cases (HH\,211, HH\,25MMS,
  V-380 Ori\,NE and HH\,212) emission was also detected at positions
  away from the center, and was stronger than that observed at the
  centre position.
  
  SiO abundances of 10$^{-8}$ to 8$\times$10$^{-7}$ are derived from
  LTE analysis. For 2 sources we have additional transitions which we
  use to conduct statistical equilibrium modeling to estimate the gas
  density in the SiO-emitting regions. For HH\,25MMS these results
  suggest that the SiO emission arises in a higher-density region than
  the methanol previously observed. We find that the most likely
  explanation for the preferential detection of SiO emission towards
  class 0 sources is the greater density of those environments,
  reinforced by higher shock velocities. We conclude that while not
  all class 0 sources exhibit SiO emission, SiO emission is a good
  signpost for the presence of class 0 sources.

\end{abstract}


\section{Introduction}

In principle SiO has great potential as an efficient shock tracer, a
millimetre-wavelength counterpart to the near-infrared lines of
H$_2$. Its abundance in quiescent gas within cloud cores is $<
10^{-12}$ relative to molecular hydrogen (Irvine, Goldsmith \&
Hjalmarson 1987; Ziurys, Friberg \& Irvine 1989), which can rise to as
high as $10^{-6}$ in outflows (Mart\'{\i}n-Pintado et al. 1992). The
formation of SiO in post-shock gas probably takes place as a
consequence of grain sputtering or grain-grain collisions releasing
Si-bearing material into the gas phase which reacts rapidly with
O-bearing species (principally O$_2$ and OH) to form SiO (Schilke et
al. 1997; Caselli, Hartquist \& Havnes 1997).

A number of outflow sources have been detected in SiO emission
(e.g. Mart\'{\i}n-Pintado, Bachiller \& Fuente 1992) but it is only
recently that extensive surveys of outflows have been carried out and
published. Harju et al. (1998) observed the locations of water and/or
OH masers, with the result that their sample is dominated by massive
young stellar objects (YSOs). Furthermore they only observed at the
maser position in each source although they were able to obtain high
sensitivity spectra. The sample of Codella, Bachiller \& Reipurth
(1999) includes YSOs with luminosities in the range 1 to 10$^4$
L$_\odot$, and they made the surprise discovery that a number of
sources show evidence for high- and low-velocity SiO components with
different spatial distributions and linewidths. Furthermore, the
abundance of SiO was different in each component, with the
high-velocity SiO two orders of magnitude more abundant than the
low-velocity SiO.  In addition, the low-velocity component was
typically only seen in the lower-lying transitions of SiO, probably
indicating that the low-velocity gas is also of lower density and/or
temperature.

In an effort to understand the conditions under which SiO emission is
observed, we have conducted a survey of the $J$=5\too 4 transition of
SiO towards a sample of 25 low-luminosity YSOs, made at the James
Clerk Maxwell Telescope (JCMT) on Mauna Kea, Hawaii. The following
section describes the observations, with subsequent sections
presenting the results, analysis and discussion of the data.

\section{Sample selection and observations}

The sources were selected from the list of Fukui et al. (1993) and/or
on the basis that comparable H$_2$ and CO observations existed.  While
the sample mostly includes low-luminosity class 0 protostars, a number
of class I and class II sources (as well as objects of somewhat
greater luminosity) are also included to test if the age and
luminosity of the driving source is relevant in the production of
observable quantities of SiO. Source details are listed in Table
\ref{obs}. 

The observations were made over several nights in 1995 October
and 1997 September, November and December using the facility receiver
A2 which was tuned to observe the 5\too 4 SiO transition at 217.1050
GHz. The backend was the digital autocorrelation spectrometer (DAS),
using a bandwidth of 500 MHz and an initial spectral resolution of 378
kHz, although the spectra were binned for the reduction and analysis
to provide a velocity resolution of 4.2 \kms\ at 217 GHz. Pointing
sources were Uranus, Saturn, 3C\,111, W3\,(OH), NGC\,7538-IRS1, NGC
2071-IR, and OMC-1. Pointing was good to 2 or 3 arcsec.
NGC\,7538-IRS1, NGC 2071-IR and OMC-1 were also used for calibration
checks by comparing with standard spectra. Position-switching was
employed throughout against a blank region of sky 300--1200 arcsec
away.

At 217 GHz, the JCMT beam has a full-width at half-maximum (FWHM) of
22 arcsec. Table \ref{obs} lists the important details for the sources
observed. Line intensities are reported as \tastar, which have been
converted to main-beam brightness temperatures, $T_{\rm
MB}$=\tastar/$\eta_{\rm MB}$ (where the main beam efficiency,
$\eta_{\rm MB}$, is 0.71 at 217 GHz), for the purposes of the
computations in Section 4.

Additional observations were made of several sources which were either
not observed in the previous runs or suffered from poor noise levels,
this time with the A3i receiver on 2000 September 4. The backend and
switching setups were identical to the previous runs. Pointing and
calibration were checked on W3\,(OH), OMC-1 and NGC\,7538-IRS1, and
were good to within 2 arcsec and 10 per cent respectively. The main
beam efficiency for these observations is 0.7.

Each source was observed at the central position, and in all cases
except two, either a grid of points was observed or a strip-map made
along the flow axis to cover the extent of the known CO flow or
H$_2$ jet. In most cases this amounted to observing 60 to 200 arcsec
away from the source position depending on the size of the flow. If
SiO emission was detected along the flow axis then in some cases we
observed off-axis positions to determine the extent of the SiO
emission (time permitting).

Only L1448-mm and NGC\,2071 were observed at a single position. Since
SiO has been mapped in both of these sources previously we did not
extend our mapping further. RNO\,43-mm was observed at a number of
positions covering the extensive CO flow (Bence, Richer \& Padman
1996). The innermost portion of the flow between the positions n1 and
s1 (Bence et al.\ 1996) was mapped as far as 150 arcsec from the
source, with further mapping carried out at the CO peaks n2, n4, n5
and n6 (Bence et al.), some 200--400 arcsec further north-east. Two
more sources were covered in our mapping of other sources: SSV\,59 was
observed while making the map of HH\,25MMS and NGC\,2024-FIR6 was
observed while mapping the NGC\,2024-FIR5 outflow. The details for all
the other sources are given in Table \ref{maps}.

\section{Results}

\subsection{HH\,211-mm}

HH\,211 is an excellent example of a Class 0 source with a molecular
hydrogen jet, driving a highly-collimated CO outflow (Gueth \&
Guilloteau 1999) perpendicular to a circumstellar disk. Knots of
shocked molecular hydrogen emission are observed in a jet-like
configuration (McCaughrean, Rayner \& Zinnecker 1994). HH\,211
exhibits the brightest 5\too 4 emission in our sample, with a peak
antenna temperature of 0.83 K in the red lobe of the outflow. The
bright line and the dual-peaked nature of the line profile led to a
successful follow up study with the VLA (Chandler \& Richer 2001). The
1\too 0 emission extends for approximately 20 arcsec either side of
the driving source and shows a velocity gradient along the jet axis
with the highest velocity emission farthest from the source. In the
transverse direction, the SiO emission extends for only an arcsec
either side of the flow axis.  Chandler \& Richer also show that the
5\too 4 emission traces the same kinematic components as the 1\too 0.

\begin{figure}
\figurenum{1}
\epsscale{0.45}
\plotone{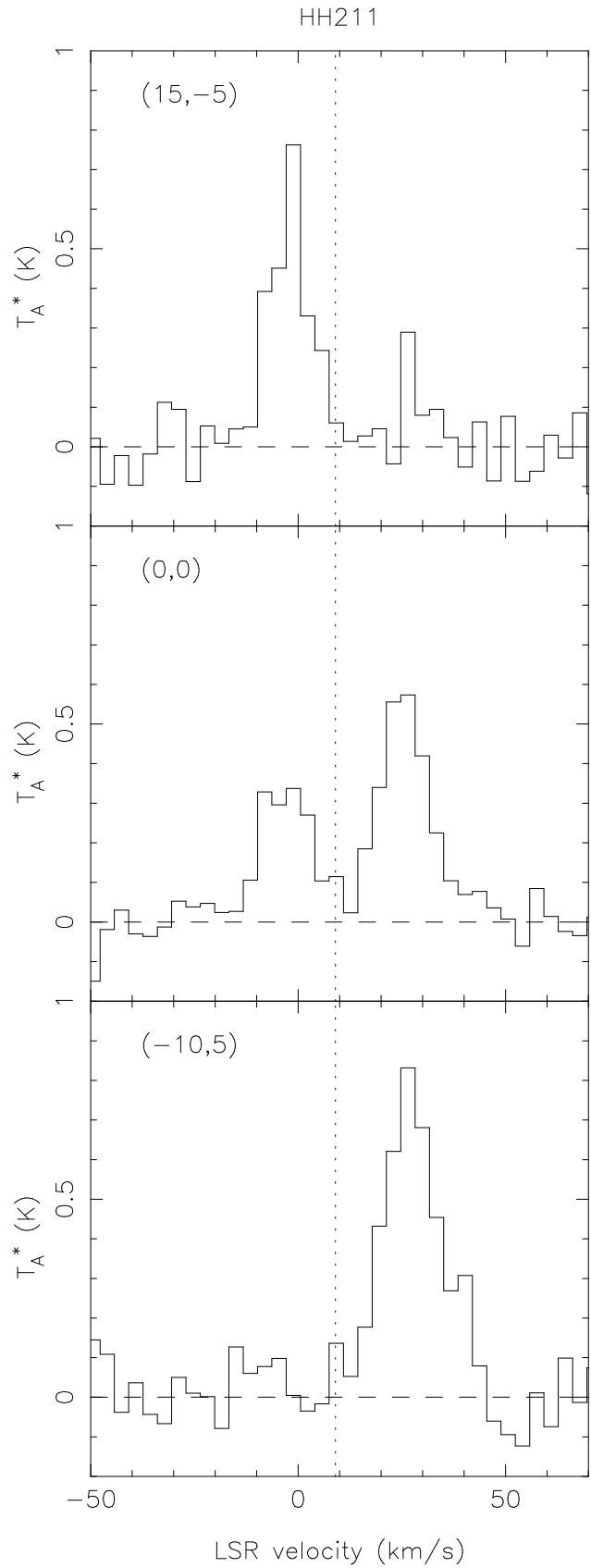}
\caption{SiO 5\too 4 spectra towards three positions in the HH\,211
  outflow. Top: from the blue lobe; Center: at the source position;
  Bottom: red lobe. The vertical dotted line marks the LSR velocity. }
\label{hh211spec}
\end{figure}

Nisini et al.\ (2002) show an image of the red- and blueshifted lobes
of the outflow as seen in SiO 5\too 4.  Each lobe peaks symmetrically
either side of the driving source and is unresolved in the 22-arcsec
beam of the JCMT.  The appearance is very similar to single-dish
observations of the CO outflow (McCaughrean et al.\ 1994). Example
spectra towards each lobe and the driving source are shown in Figure
\ref{hh211spec}. The two velocity components of the outflow are
clearly spectrally resolved. There is virtually no SiO emission at the
systemic velocity: the bulk of the emission peaks 10--20 \kms\ from
the rest velocity. The velocity extent of each lobe is approximately
20 \kms.  Gaussian fits to spectra at the peak of each lobe give
linewidths of $\sim$10--16 \kms\ (blue lobe) and 15 \kms\ (red lobe).
 
\subsection{HH\,25MMS}

In our survey, HH\,25MMS exhibits the most spatially-extended SiO
emission of all the targets in Table \ref{obs}. HH\,25MMS is a
deeply-embedded low-luminosity (6 L$_\odot$) class 0 protostar in the
L\,1630 molecular cloud (Gibb \& Davis 1998, hereafter GD98). A
compact but highly-collimated CO flow which probably lies close to the
plane of the sky is centered on the position of a weak radio continuum
source (GD98; Bontemps, Andr\'e \& Ward-Thompson 1995). In addition a
number of other molecules are seen in the outflow, in particular CS,
CH$_3$OH and H$_2$CO (GD98).

\begin{figure}
\figurenum{2}
\epsscale{1.0}
\plotone{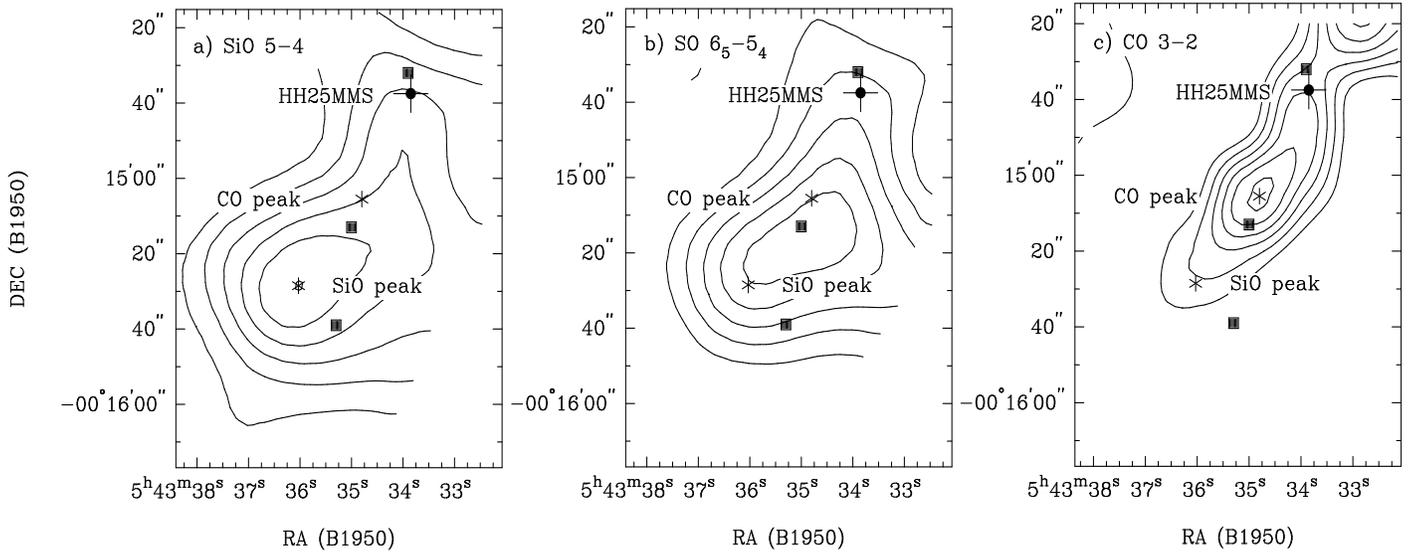}
\caption{a) SiO 5\too 4 integrated intensity in the range 8 to 15
  \kms. b) SO 6$_5$\too 5$_4$ integrated intensity in the range 8 to
  12 \kms). c) Redshifted CO 3\too 2 emission from the HH\,25MMS
  outflow for comparison (Fig. 1a from GD98). In all three panels, the
  position of HH\,25MMS is marked by a cross and filled circle, the
  knots comprising HH25 as open squares and the SiO and CO peaks by
  asterisks. }
\label{hh25mms}
\end{figure}

The SiO emission is undoubtedly associated with the CO outflow (see
Fig. \ref{hh25mms}) although there are significant differences in
their respective distribution. The respective CO and SiO peaks are
separated by $\sim$35 arcsec with the SiO peaking at the tip of the CO
flow, offset by (36,--57) arcsec from HH\,25MMS. The CO peak at
(14,--29) is not distinguished in the SiO map. The H$_2$ knots HH25C
and HH25D (Davis et al. 1997a) appear to be closely associated with the
SiO, evidence which supports the formation of SiO upstream from bow
shocks.  However, it is not possible to directly associate the H$_2$
emission features with any particular SiO feature. Higher resolution
observations will undoubtedly help to clarify this relation.

The SiO 5\too 4 observations of HH\,25MMS were tuned to obtain the
$6_5$\too $5_4$ line of SO (in the upper sideband at 219.9494 GHz)
within the same passband and is clearly detected in HH\,25MMS.
Figure~\ref{hh25mms}b shows the distribution of SO, which is largely
similar to the SiO. The peak SO emission is brighter than the SiO
(\tastar $\sim$ 0.9 K towards the peak position compared with 0.5 K
for the SiO), but generally narrower (linewidths typically 4 \kms\ 
compared with 9--12 \kms\ for the SiO). The SO emission has a similar
distribution to the methanol seen by GD98, suggesting that the action
of the outflow is directly responsible for driving a series of
chemical reactions not otherwise possible in the ambient gas.

Since the signal-to-noise ratio is relatively low, little kinematic
information can be obtained from velocity channel maps. However, it is
still possible to distinguish two velocity components to the SiO
emission, one redshifted relative to the line center, the other
blueshifted. The dominant component is the redshifted which follows
the CO lobe very closely (Fig. \ref{hh25mms}c), and peaks at the very
tip of the flow. The blueshifted component is weaker and peaks
approximately 10 arcsec west of HH\,25D, giving the SiO
image in Fig.~\ref{hh25mms} its westward skew. The mean SiO linewidth
derived from gaussian fits is $\sim$7.5 \kms, although many of the
spectra are not symmetric.

\begin{figure}
\figurenum{3}
\epsscale{0.45}
\plotone{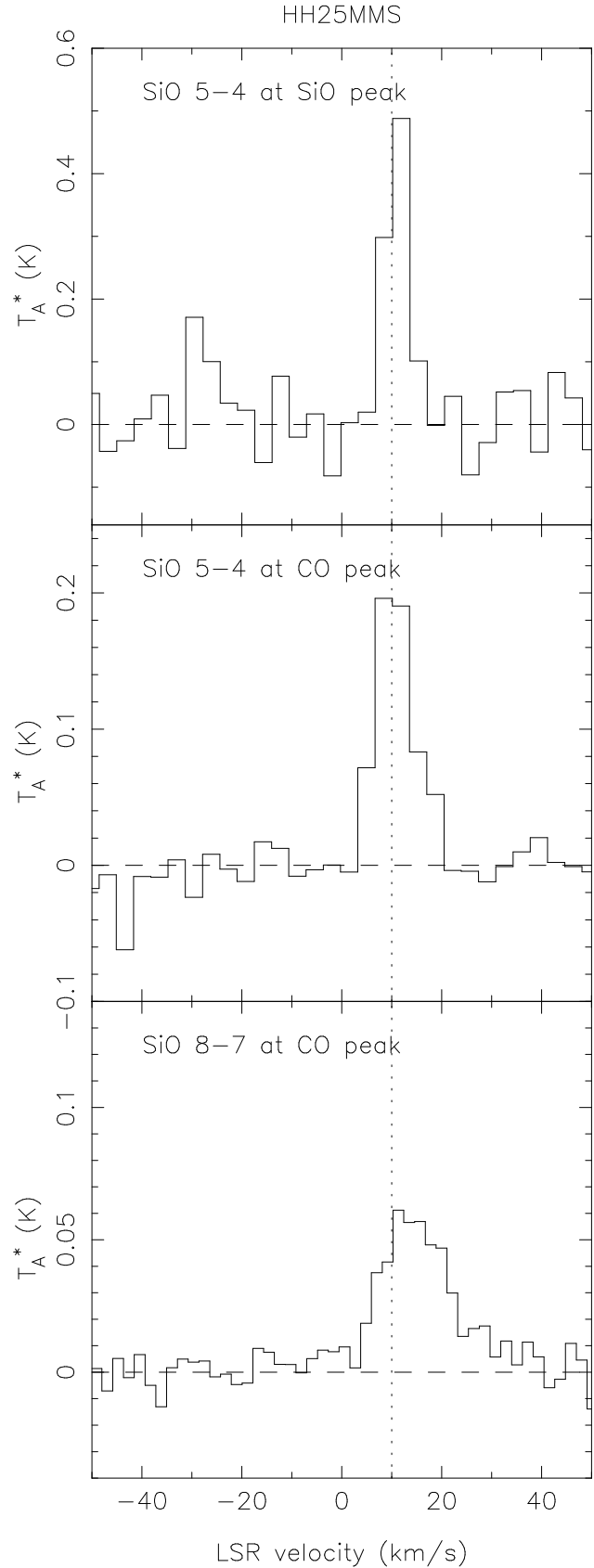}
\caption{Top: SiO 5\too 4 spectrum towards the SiO peak.
  Center: SiO 5\too 4 spectrum towards the CO peak. Bottom: SiO 8\too
  7 spectrum towards the CO peak. The noise levels are 52, 20 and 6 mK
  per 2.5-MHz channel respectively. The vertical dotted line
  marks the LSR velocity. 
}
\label{hh25spec}
\end{figure}

In addition to the 5\too 4 data, we have a small map of the 7\too 6
line (GD98) and a single long integration 8\too 7 spectrum.
Unfortunately, the 7\too 6 map does not extend as far as the tip of
the outflow. The 7\too 6 spectrum shown in GD98 is from the CO peak.
The width of the 7\too 6 line is 8.5 \kms, and it peaks at a velocity
redshifted by $\sim$5 \kms\ relative to the systemic velocity. The
single 8\too 7 spectrum is also from the CO peak, shown in Fig.
\ref{hh25spec}c.  Like the 7\too 6 line, the 8\too 7 line at the CO
peak is redshifted, centered at +5 \kms\ relative to the systemic
velocity, with a FWHM of 9.9 \kms.

\subsection{HH\,212}

\begin{figure}
\figurenum{4}
\plotone{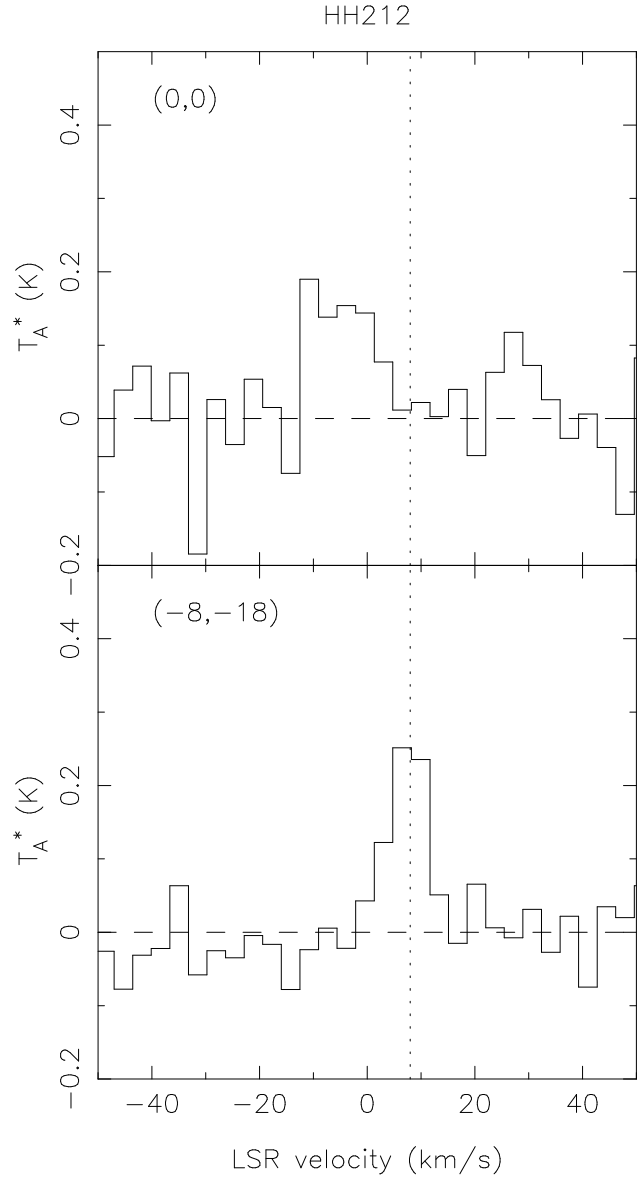}
\caption{Top: SiO 5\too 4 line towards the central position in
  HH\,212, and Bottom: towards (--8,--18) arcsec offset. The line
  shown peaks at --6.0 \kms and has a width of $\sim$16 \kms. Note
  that in the top panel there may be redshifted counterpart to the
  highly blueshifted line, peaking close to 28 \kms. However, this
  peak is only detected at the 2-$\sigma$ level. The vertical dotted
  line marks the LSR velocity.}
\label{hh212}
\end{figure}

HH\,212 is also a low-luminosity class 0 protostar associated with the
IRAS source 05413--0104 which drives a bipolar H$_2$ jet (Zinnecker,
McCaughrean \& Rayner 1998). HH\,212 is a water maser source and
high-resolution observations have revealed the proper motions of the
maser spots in the southern outflow lobe (Claussen et al. 1998). The
H$_2$ jet also drives a well-collimated CO flow (Lee et al. 2000). As
seen in Fig. \ref{hh212}, the spectra at the center position and
(--8,--18) offset show clear detections. The line at the center
position can be fitted by a gaussian of peak \tastar\ = 0.17 K with a
FWHM of 15.8 \kms, centered at --12.8 \kms\ relative to the rest
velocity of the cloud. The line at the (--8,--18) position is not
significantly shifted with respect to the ambient cloud, and is
narrower (8.3 \kms).

\subsection{L\,1448-mm}

L\,1448-mm (also known as L\,1448C) drives a remarkable
highly-collimated bipolar outflow, and was the first in which
kinematically-distinct molecular clumps (`bullets') were detected
(Bachiller et al. 1990) at velocities up to 70 \kms\ from the systemic
velocity of the cloud core. SiO 2\too 1 emission was first detected by
Bachiller, Mart\'{\i}n-Pintado \& Fuente(1991) and mapped in more
detail by Dutrey, Guilloteau \& Bachiller (1997), so the detection of
the 5\too 4 line was not unexpected. Chernin (1995) discovered water
maser emission within the confines of the flow. 

\begin{figure}
\figurenum{5}
\plotone{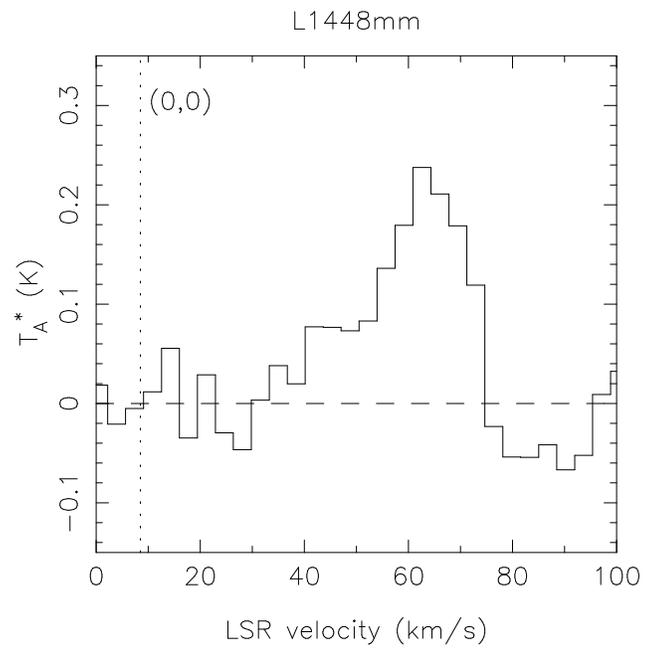}
\caption{SiO 5\too 4 spectra towards the L\,1448 outflow. The vertical
  dotted line marks the LSR velocity. }
\label{l1448spec}
\end{figure}

A single spectrum was recorded at the position of the millimetre
source and is shown in Figure \ref{l1448spec}. The SiO 5\too 4 line is
detected with a peak antenna temperature (\tastar) of 0.25 K and a
gaussian fit to the line yields a linewidth of 15 \kms\ centered at an
LSR velocity of 64 \kms. The most notable aspect to this detection is
its large redshift of 55 \kms\ with respect to the systemic velocity.
This is in good agreement with the interferometer maps of Dutrey et
al. (1997) which show an SiO peak close to L\,1448-mm over an LSR
velocity range of 50--75 \kms.

\subsection{V380 Ori-NE}

\begin{figure}
\figurenum{6}
\plotone{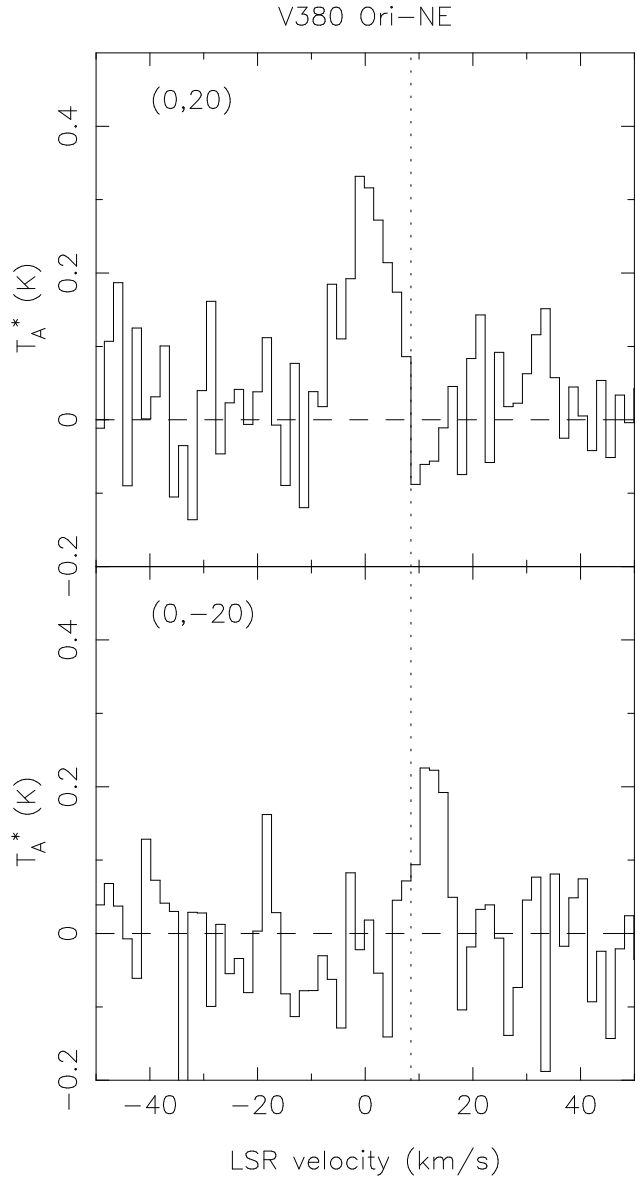}
\caption{SiO 5\too 4 spectra towards the positions B1 and R1 (in the
  nomenclature of Davis et al.\ 2000) in the outflow from V380 Ori-NE.
  Top: 20 arcsec north of center (B1), and Bottom: 20 arcsec south
  (R1). Note that the line towards B1 is blueshifted by nearly 10
  \kms\ relative to the systemic velocity, while that towards R1 is
  only slightly redshifted (+2.2 \kms). The vertical dotted line marks
  the LSR velocity. The spectral resolution is 1.25 MHz in these
  figures.}
\label{v380}
\end{figure}

The outflow from V380 Ori-NE is orientated roughly north-south and
displays a high degree of collimation (Davis et al. 2000). Although
classified as a class I source by Zavagno et al. (1997), it shares
some characteristics with class 0 sources (such as a highly collimated
outflow) and may be a transitional object. The SiO spectra in Fig.
\ref{v380} are from the positions of the CO peaks B1 and R1 (in the
nomenclature of Davis et al. 2000). Like the CO 4\too 3 emission, the
SiO line towards B1 is brighter than towards R1. The SiO detection at
B1 is also broader than that at R1 (9.9 \kms\ compared with 6.3 \kms).
These positions are also where the H$_2$ jet curves round, possibly in
an interaction with the core in which the driving source is still
embedded, although the jet bending may be due to precession of the
central source (Davis et al. 2000). However, the observation of SiO
emission at these same positions supports the interaction
interpretation.

\subsection{NGC\,2071}

NGC\,2071 is a well-known intermediate-luminosity ($\sim$500
L$_\odot$) star-forming region which has long been known to exhibit
SiO emission (Chernin \& Masson 1993; Garay, Mardones \&
Rodr\'{\i}guez 2000) in its CO outflow (Moriarty-Schieven, Snell \&
Hughes 1989). NGC\,2071 also houses a cluster of water masers which
appear to delineate a compact jet-like feature in the outflow
(Torrelles et al. 1998). The 2\too 1 and 3\too 2 SiO emission follows
the outflow, tracing two lobes which peak 40 arcsec either side of the
exciting source.

A single SiO 5\too 4 spectrum was recorded at the position of the
outflow driving source, NGC\,2071-IR. The line is detected at an LSR
velocity of 2.5 \kms, blueshifted with respect to the systemic
velocity. A gaussian fit yields a peak \tastar\ of 0.25 K and a FWHM
of 7.1 \kms.

\subsection{NGC\,2024-FIR6}

The very compact and high-velocity outflow from NGC\,2024-FIR6 was
discovered by Richer (1990). FIR6 is the site of a dust continuum peak
(Mezger et al. 1988; Visser et al. 1998) and houses a moderately
luminous young stellar object, estimated at 3000 L$_\odot$, and is the
site of a water maser. Our SiO spectrum shows a tentative detection
(3$\sigma$ peak, 4$\sigma$ in integrated intensity) at an LSR velocity
of --7.2 \kms. This suggests an association with the very
high-velocity blue lobe of the FIR6 outflow.

\subsection{Non-detections}
\label{nondet}

Of the remaining sources, none was detected at a level of 3$\sigma$ or
greater.  Table \ref{obs} lists the peak integrated intensities and
main-beam brightness temperatures for the detected sources, while
upper limits are given for the non-detections. The upper limits to the
brightness temperature are derived from the mean noise level per
2.5-MHz channel in the maps made of each source. The upper limits to
the integrated intensity are estimated by integrating the noise per
channel over a velocity interval of 12 \kms, which represents the mean
velocity extent of the detected sources.

Some of our non-detections are surprising given that a significant
number have well-defined outflows and jets. For example, HH\,7--11 has
a dense component to its outflow, seen in CS and HCO$^+$ (Rudolph et
al.  2001; Rudolph \& Welch 1988). The extensive SiO map made by
Codella et al. (1999) verifies that the main HH\,7--11 flow has no SiO
emission along the axis, but does appear to bounded by a cavity
defined by emission from the lower energy 2\too 1 line. Codella et
al.\ did not detect this component in the 5\too 4 line. However, the
nearby source SVS13B (which we did not cover in our mapping) does show
strong 5\too 4 emission while the 2\too 1 traces a jet-like outflow
(Bachiller et al. 1998).

Cep E was detected by Lefloch, Eisl\"offel \& Lazareff (1996) in the
3\too 2 line of SiO with a velocity spread of 30 \kms. The
non-detection of the 5\too 4 line indicates that the density within
the JCMT beam is probably not high enough to sufficiently excite this
line.

Other notable non-detections are HH\,24MMS, NGC\,2024-FIR5 (30 arcsec
north-west of FIR6 and the source of a highly-collimated CO flow:
Richer, Hills \& Padman 1992), RNO\,43 (with its 5-pc long CO flow:
Bence et al.\ 1996) and HH\,111 which shows evidence of molecular
`bullets' (like L1448-mm: see above) in addition to a well-defined
outflow cavity (Nagar et al. 1997; Hatchell, Fuller \& Ladd
1999). Possible reasons for the non-detection of SiO towards these
sources are discussed in section 5.

\section{Analysis}

In this section we calculate SiO column densities from the observed
spectra under the assumption that the lines are optically thin, and
that the level populations are in local thermodynamic equilibrium
(LTE) which can be characterized by a single temperature. The SiO
emission probably arises from regions much smaller than the JCMT beam
and so we use main-beam brightness temperatures in our calculations.
We also carry out radiative transfer analysis of the emission from
HH\,25MMS and HH\,212 and summarize this in a following subsection.

\subsection{LTE column densities and abundances}
\label{lte}

The calculation of beam-averaged SiO column densities employed the
following formula, adapted from Irvine et al. (1987):
\begin{equation}
N_{\rm SiO} = 2.5 \times 10^{12} \int T_{\rm mb}\,dv
\end{equation}
The column density, $N_{\rm SiO}$, is in cm$^{-2}$ and the
integrated intensity, $\int T_{\rm mb}\,dv$, is in K~\kms. A dipole
moment of 3.1 D has been assumed (Ziurys et al. 1989).
The fact that we have only a single transition for most sources and
that the SiO excitation temperature is not known means that the column
densities we derive are strict lower limits if we assume an excitation
temperature equal to the energy of the upper level (Macdonald et
al. 1996). The $J$=5 level lies 31 K above the ground state; at this
temperature, the fraction of SiO molecules in the $J$=5 level is
0.133. Table \ref{nsio} lists the SiO column densities for sources
with detected SiO 5\too 4 emission.

For comparison, CO column densities have also been calculated, and
here the choice of velocity range over which to integrate has been
chosen to match that of the SiO as closely as possible. In the case of
HH\,212, for example, both the CO and SiO display redshifted emission
and thus it is simply a case of integrating over the same velocity
range for each line. For HH\,25MMS the SiO emission tends to lie at
the systemic velocity while the CO shows both red- and blueshifted
components. Since the SiO is clearly spatially associated with the
high-velocity gas, it would be incorrect to integrate the CO over the
line center. Our approach has been to calculate the ratio of
red-shifted SiO to red-shifted high-velocity CO. Finally a
beam-averaged abundance relative to molecular hydrogen is also
presented in Table \ref{nsio}, where we have assumed that the CO/H$_2$
ratio is 10$^{-4}$.

The ratios presented in Table \ref{nsio} should clearly be treated
with some degree of caution, especially as both numerator and
denominator are lower limits. A major source of uncertainty is the
unknown beam-filling factor of the SiO emission, likely to be
significantly lower than the CO (e.g. Chandler \& Richer 2001; Dutrey
et al. 1997).  Allowing for this would increase the SiO column density
relative to the CO and thus raise the derived SiO abundance. Another
source of uncertainty in the calculations is the choice of excitation
temperature. For example, Chandler \& Richer (2001) derived values of
150--200 K for HH\,211.  The fractional population of the respective
upper levels for both CO and SiO changes by less than a factor of 3 in
the range 30 to 200 K. Assuming the CO and SiO emission arises from
the same region leads to higher column densities for each but
unchanged abundances. In reality the CO emission is probably more
widespread (and thus has a lower mean excitation temperture) than the
SiO, so the CO column density {\em within the SiO-emitting region} is
likely to be lower than listed in Table \ref{nsio}. Therefore the SiO
abundance is probably higher than given in Table \ref{nsio}.

\subsection{Radiative transfer modeling}

\begin{figure}
\figurenum{9}
\epsscale{0.9}
\plotone{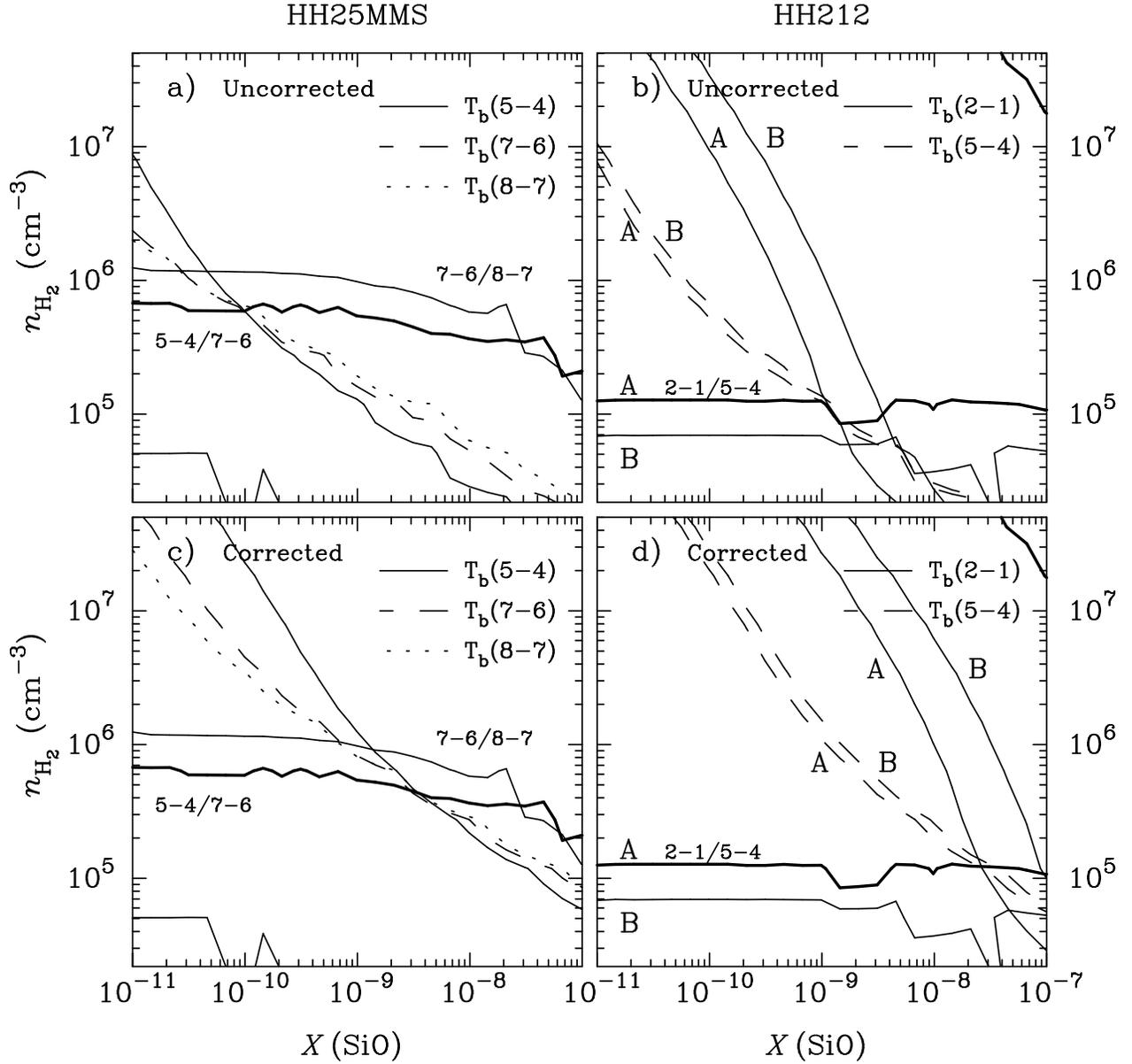}
\caption{Plot of LVG solutions for a kinetic temperature of 100 K for
a) and c) HH\,25MMS and b) and d) HH\,212. In all panels, the contours
showing the brightness temperatures run top-left to bottom-right,
while the observed line ratios run roughly horizontally across the
figure. The top panels show the observed brightness temperatures
(Uncorrected), while the lower ones show the brightness temperatures
(Corrected) when a beam-dilution correction is applied (multiplied by
20). For the HH\,212 figures, the results are given for two positions
marked A and B. Position A corresponds to the central position while B
corresponds to ($-$8,$-$18).  See text for full details. }
\label{lvgfig}
\end{figure}

In addition to the LTE analysis, for HH\,25MMS
and HH\,212 we have data in more than one transition and thus have
attempted to further constrain the properties of the shocked gas with
a statistical equilibrium code employing the large velocity gradient
(LVG) approximation (Goldreich \& Kwan 1974). Collision rates were
taken from Turner et al. (1992). The parameter space covered by the
modeling extends from a density of 10$^4$ cm$^{-3}$ to 10$^9$
cm$^{-3}$, a temperature of 20 to 150 K and an abundance of SiO from
10$^{-11}$ to $10^{-7}$ relative to H$_2$. The velocity gradient was
assumed to be the FWHM linewidth of the SiO line divided by the beam
diameter. Since the emitting regions are likely to be much smaller
than the beam, fitting the line brightnesses produced solutions with
low beam-averaged abundances, lower than derived from the LTE analysis
in the previous section. Since the line ratios constrain the density
fairly well while being only weakly dependent on temperature, the only
parameter which is likely to be in error is the SiO abundance.  The
solutions for each source are described in more detail below and are
presented graphically in Fig.~\ref{lvgfig}.

\subsubsection{HH\,25MMS}

At the location of the CO peak in HH\,25MMS we have long-integration
spectra of the 5\too 4 and 8\too 7 transitions, plus a
lower-sensitivity 7\too 6 spectrum (see Fig. \ref{hh25spec}). Table
\ref{lvg} lists the relevant parameters of these lines. 

The results are not particularly well constrained and no single
solution could be obtained. The range of acceptable solutions required
a kinetic temperature 50--100 K, an SiO abundance of 2--20$\times
10^{-11}$ and a density of order $10^6$ cm$^{-3}$. Higher
abundance solutions were ruled out on the basis that the line ratio
required a higher density solution compared with the observed line
intensities. Fig. \ref{lvgfig}a shows a typical solution for a
temperature of 100 K.

It is interesting to compare these results with those derived for
methanol at the same position by GD98. In this case, a temperature of
about 60 K gave the best fit as well as a lower density than for the
SiO (1--4$\times$10$^5$ cm$^{-3}$). This perhaps begins to provide a
picture of how the various molecules are distributed in shocked gas:
the SiO lies in the higher density gas, immediately post-shock while
molecules such as methanol exist in a lower density
environment. Higher-resolution observations are critical to testing
this hypothesis.

\subsubsection{HH\,212}

Chapman et al.\ (2002) report the detection of SiO 2\too 1 emission
close to the location of driving source of the HH\,212 jet. They
detect two velocity components, in good agreement with our 5\too 4
detections shown in Figure \ref{hh212} suggesting that they are
tracing the same components that we detect in the 5\too 4 line. Table
\ref{lvg} summarizes the results of the LVG modeling. Similar
solutions are found for each position, although the temperature is
slightly better constrained for the SW position at (--8,--18). The
density in the SiO-emitting region is of order 10$^5$ cm$^{-3}$ and
the beam-averaged SiO abundance is estimated to be a few times
10$^{-9}$. Fig. \ref{lvgfig}b shows a typical solution for a
temperature of 100 K.

\subsection{LVG modeling discussion}

Interpreting the results of the LVG modeling is not necessarily
straightforward, so here we elaborate on the meaning of the results.
The LVG approximation assumes a constant velocity gradient across the
clump of molecular gas. In addition, in our modeling, the clump has
constant kinetic temperature and density. For a set of inputs (line
brightness temperatures and ratios) the LVG code predicts the
brightness temperature of each SiO transition for a range of
densities, temperatures and abundances. Since the temperature and
density are assumed to be constant, the same is true for the
abundance.

In practice the effective output abundance, $X$, is the abundance
divided by the velocity gradient, $dv/dr$, i.e. $X/(dv/dr)$ (see
Goldreich \& Kwan 1974).  Thus models may be run for one value of
$dv/dr$ and solutions for the abundance may be scaled to other
possible values of $dv/dr$.  Furthermore, if the emission arises from
a region much smaller than the beam, then the true abundance is higher
since $X/(dv/dr)$ is constant. Thus the abundances we derive from LVG
modeling are lower limits. In the optically thin limit, a
high-density/low-abundance solution is qualitatively the same as a
low-density/high-abundance solution.

Since there are three parameters being solved for, three input values are
necessary. Ideally three independent values are desireable (i.e.\ three
different transitions), but in our case we use the two line brightness
temperatures and their ratio.

The solutions given in Table \ref{lvg} yield abundances for SiO that
are lower than the LTE values derived in Section \ref{lte}. While the
observed line ratios do not rule out higher abundances, the observed
line {\em intensities} do: higher abundance solutions predict higher
line intensities. Therefore, if the true abundance of SiO is higher
than that predicted by the LVG modeling, then the observed line
intensities must be low as a result of beam dilution.  In section
\ref{beamfilling} below, we estimate that the filling factor for the
5\too 4 emission probably lies between 0.01 and 0.2.  If we take the
(geometric) mean and correct the observed line intensities accordingly
(assuming that all the lines are equally affected) then the LVG models
predict solutions with a higher abundance than without the correction
applied. This is shown by the `Corrected' curves in Fig.~\ref{lvgfig}c
and d.  The increase in the abundance is roughly proportional to the
reciprocal of the filling factor since the SiO emission in the models
is optically thin.

\section{Discussion}

In this section we will discuss our results and investigate possible
reasons for the lack of widespread SiO 5\too 4 emission towards our
sample of outflow sources. Note that this discussion is biased toward
understanding the factors influencing only the detection of the 5\too
4 line.  As shown by Codella et al.\ (1999), lower-lying transitions
may trace different excitation conditions to which the 5\too 4 line is
not sensitive.

To recap on the results presented above, of 25 outflow sources
observed, SiO 5\too 4 emission was detected in only 7, a 28 per cent
detection rate. From these data there is weak evidence for a
correlation between the evolutionary status of the outflow driving
source and SiO emission: 5 out of 7 of the detected sources satisfy
the class 0 criteria and are presumably relatively young. The results
from the LVG modeling suggest that the molecular hydrogen density in
the SiO-emitting regions is of order 10$^5$--10$^6$ cm$^{-3}$.

\subsection{Which sources show SiO emission?}

Here we examine a number of potential correlations between source
properties and SiO emission in order to assess whether or not these
parameters play a role in determining the strength of SiO emission.
The first correlations to examine are the obvious ones involving
outflow velocity and power. Values for the outflow parameters have
been taken from the literature where possible. In the cases where none
has been published, then we have assumed that the mean velocity is
half the maximum velocity and the mean radius is half the maximum
radius of the outflow (where the lobes are of unequal size, an average
of those values was used). In addition we have calculated the outflow
parameters in the following way. The dynamical timescale for the
outflow, $t_{\rm dyn} = 9.8\times 10^5 r/v$ years, where the radius,
$r$, is in pc and the velocity $v$ is in \kms, the momentum flux,
$\dot{p} = Mv/t_{\rm dyn}$ M$_\odot$\,\kms\,yr$^{-1}$, where $M$ is
the outflow mass in M$_\odot$, and the mechanical luminosity $L_{\rm
mech} = 81 \dot{p} v$~L$_\odot$.

In our attempt to understand the processes which give rise to
detectable SiO emission, we examine possible correlations between the
observed SiO emission and a number of source properties. Since SiO is
formed in shocked gas, it is reasonable to search for correlations
between the SiO emission and outflow properties, such as the velocity,
outflow force and mechanical luminosity. These properties all likely
reflect the driving source (stellar) luminosity, so we do not
separately plot this quantity. Since the SiO lines require high
densities to excite, another property which may influence the SiO
emission is the density of the medium through which the outflow is
propagating.

\begin{figure}
\figurenum{10}
\epsscale{0.9}
\plotone{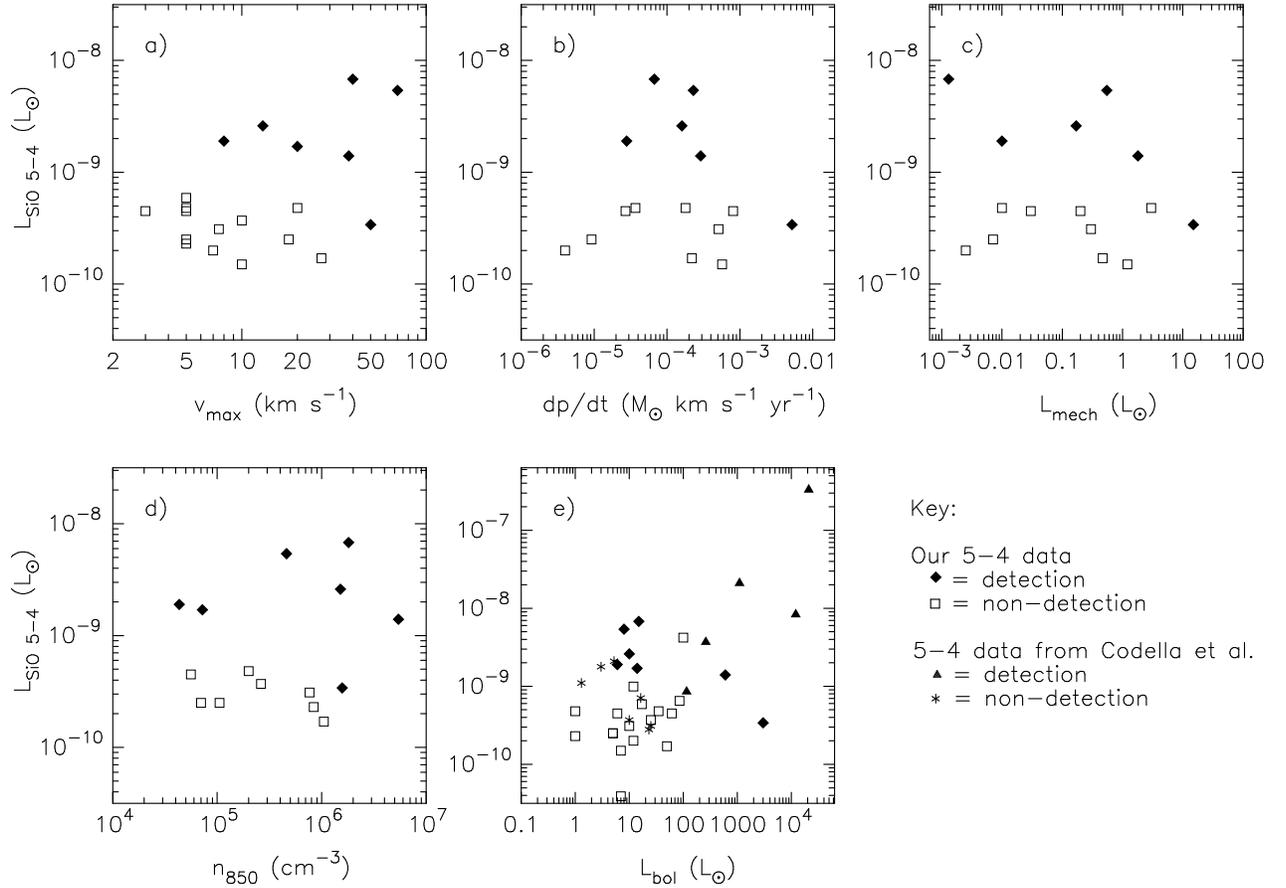}
\caption{Plot of SiO $J$=5\too 4 luminosity as a function of source properties
  a) Maximum velocity, b) outflow force, c) mechanical luminosity and
  d) submillimetre density. SiO detections are plotted as filled
  diamonds; upper limits are denoted by open squares.  The
  uncertainties in the outflow parameters are typically a factor of
  2--3, somewhat less than the spread in values. The uncertainty in
  the submillimetre density is a factor of $\sim$2. Panel (e) plots
  the SiO $J$=5\too 4 luminosity as a function of source bolometric
  luminosity, and includes data from Codella et al.\ (1999) as 
  triangles.}
\label{sioplots}
\end{figure}

Figure \ref{sioplots} shows the SiO 5\too 4 luminosity (in units of
L$_\odot$) plotted against five source properties: the maximum
velocity in the outflow, $V_{\rm max}$, the outflow force, mechanical
luminosity, number density of H$_2$  derived from the submillimetre
continuum emission (or the `submillimetre density': see section
\ref{density} below) and source bolometric luminosity. Note that we
only include data for the sources which lie in the distance range 330
to 450 pc (see section \ref{coverage} below). In estimating the
density, we have assumed that the emission comes from a region 15
arcsec in diameter (equal to the JCMT beam at 850 $\mu$m) and have
used published dust temperatures where available.  The precise
emissivity used is not important because we are only interested in
relative differences between the sources and we assume that the
emissivity is the same for all the sources in our sample.

No clear correlations are observed between SiO emission and any of the
plotted properties, although Fig. \ref{sioplots} does suggest that
sources with faster outflows are more likely to exhibit SiO emission,
and there is a weak trend that sources with higher ambient gas
densities are also more likely to be detected in our survey (see the
following section for more detailed discussion). Contrary to the
results of Codella et al.\ (1999; Fig.~\ref{sioplots}e) we find no
unambiguous correlation between source luminosity and strength of SiO
emission.  However, it should be noted that the Codella et al.\ sample
includes higher luminosity sources, and includes only one young,
low-mass, class 0 source (SVS13B).  Since we conclude below that SiO
is a good indicator of such objects, it is perhaps not surprising that
we do not observe any correlation.  The Codella et al.\ observation
that the more luminous sources exhibit stronger SiO emission can in
fact be explained by the same conclusion we reach, namely that the
ambient density plays some role in determining the presence of SiO
emission.

\subsection{Factors influencing SiO production and excitation}

\subsubsection{Shock velocity}

The shock velocity needed to disrupt dust grains sufficiently to
release Si from grain mantles into the gas phase is of order 25 \kms\ 
(Schilke et al. 1997). Most of our SiO detections are from outflows
with (line-of-sight) velocities exceeding 20 km\,s$^{-1}$. If
projection effects are taken into account, then it is likely that all
the SiO detected-flows have velocities greater than 20 km\,s$^{-1}$
(e.g. the HH\,25MMS flow probably lies within 20 degrees of the plane
of the sky, leading to a corrected flow velocity of at least 24 \kms:
see GD98). The majority of sources in our sample, however, have
$V_{\rm max}$ below this value.  There is no clear correlation between
SiO line strength and maximum outflow velocity although this is
probably because there are other factors which determine the intensity
of SiO emission (and these are discussed below). However, of the
outflow sources with maximum (line-of-sight) velocities greater than
20 km\,s$^{-1}$, almost half of them exhibit SiO emission. In this
sample, on average, the outflows with SiO emission have higher CO
velocities than those without (30 km\,s$^{-1}$ compared with 15
km\,s$^{-1}$). Therefore, as one might expect, the outflow velocity
does appear to play some role in determining the degree of SiO
emission.

\subsubsection{Density}
\label{density}

A high shock velocity alone is probably insufficient to lead to strong SiO
emission: the density must also be a factor. The critical density for the
excitation of the 5\too 4 line of SiO is of order 5--10$\times$10$^6$
cm$^{-3}$. (Those for the 7\too 6 and 8\too 7 lines are more than a
factor of ten higher.) 
Therefore in order for this line to be detected in our survey,
either the gas must be very hot for beam dilution not to render the
line undetectable, or the mean density across a 22-arcsec diameter
beam must be at least a few times 10$^6$ cm$^{-3}$.
Our LVG results (above) confirm that in those sources which are
detected in SiO emission, the density is indeed probably close to
this value, and in the case of HH\,25MMS that the higher transitions
are not thermalized.

Of course, the SiO emission originates in the post-shock gas, so the
densities derived above refer to this component. The compression ratio
for a C-shock is of order $\sqrt{2}$ times the magnetic Mach number
(shock speed divided by the Alfv\'en velocity: e.g.\ Spitzer 1978).
For a shock velocity of 25 km\,s$^{-1}$ and an Alfv\'en velocity of
1.5--2.5 km\,s$^{-1}$ (estimated from the linewidths of ambient gas
tracers), the compression in a C-shock is of order 15--25 (as borne
out by models of shocks: e.g.\ Wardle 1991).  Therefore, in order for
the post-shock gas to reach densities of order 5$\times$10$^6$
cm$^{-3}$, then the pre-shock density needs to be at least
2$\times$10$^5$ cm$^{-3}$, a somewhat less stringent limit but still
higher than the typical density of low-mass dense cloud cores, such as
those traced by ammonia (Benson \& Myers 1989). Therefore it seems
reasonable to expect that higher density regions are favoured for the
production of SiO 5\too 4 emission.

As a way of exploring whether the density of the {\em ambient} gas
plays a role in determining the strength of the SiO emission, we have
plotted the SiO luminosity against the submillimetre density, a
quantity designed to represent the mean ambient density within a
15-arcsec beam in Figure \ref{sioplots}.
As in the case for the maximum velocity, there is no clear trend but
on average it does appear from this crude analysis that the sources
which show SiO emission lie in regions with higher densities than
those which are not detected.

\subsubsection{Incomplete beam-filling}
\label{beamfilling}

From the LVG analysis it is evident that the SiO lines in HH\,25MMS
originate from unresolved components within a 15-arcsec beam because
the solutions for the line ratios predict higher abundances than the
line intensities. To bring the solutions into agreement we can assume
the emission fills a small fraction of the beam and scale the line
brightnesses to derive a consistent solution. Using this approach it
is possible to estimate that the lines have beam-filling factors of
order 0.01 to 0.2. These filling factors correspond to source
diameters in the range 1.5 to 7 arcsec, and are in good agreement with
estimates in other sources (e.g.\ Avery \& Chiao 1996). Such size
scales are comfortably within the range of current interferometers and
a future test would be to observe the detected outflow sources with
arcsecond resolution. The VLA study of HH\,211 by Chandler \& Richer
(2001) show that this estimate is not unreasonable.

\subsubsection{Chemistry in shocked gas}

While the abundance of SiO is enhanced by the action of the shock, the
density may be sufficiently high such that the depletion timescale
becomes shorter than the shock crossing timescale, thus `quenching'
emission from SiO by lowering the abundance. Such conditions may be
met in very dense regions and where the post-shock gas quickly cools
to 100 K or less at which point the SiO can then freeze out onto the
dust grains. From the models of Schilke et al. (1997), the shock
crossing time is calculated to be 20000 years for a shock speed of 25
km\,s$^{-1}$. The depletion timescale is $2\times 10^9/n_{\rm H_2}$
years (where $n_{\rm H_2}$ is the molecular hydrogen number density in
cm$^{-3}$). These two values are equal when the density is $10^5$
cm$^{-3}$, marginally lower than the post-shock densities inferred
above. This may account for the non-detection of SiO towards
high ambient density sources such as HH\,24MMS and NGC\,2024-FIR5.

Another mechanism for removing SiO from the gas phase (once formed) is
for it to pass through a second shock in which reactions with OH
(abundance in shocks) rapidly form SiO$_2$. Schilke et al.\ (1997)
demonstrate in their shock models that a region where SiO is the
dominant pre-shock reservoir for silicon will have a lower maximum
abundance of SiO (by up to an order of magnitude) compared with those
regions where Si is in other forms.

\subsubsection{Spatial coverage and sensitivity}
\label{coverage}

We have considered whether the non-detections in our survey result
from mapping an insufficiently large area.  Our mapping scheme
involved two approaches: a grid of points covering the outflow source
and its environment (for small outflows), and strip maps (for large
outflows) where spectra were recorded at beamwidth spacing covering
the extent of the known CO outflow.  In the latter case, if SiO
emission was detected along this axis then the mapping was extended to
cover off-axis points (time permitting).  In one case (RNO43-mm) the
flow was too large to cover completely so the mapping focused on the
region close to the driving source and the positions of the known CO
and/or H\,$\alpha$ peaks.  It is worth noting that for sources where
off-axis SiO emission was detected, emission {\em on} the outflow axis
was detected as well.  Examining the SiO detections from other
low-mass outflow sources in the literature (see
Section~\ref{low-mass}) we note that in all cases, an on-axis
strip-map with a 22-arcsec beam would have detected the SiO emission.
We therefore conclude that poor spatial coverage is not a contributing
factor to our non-detections.

We also consider whether the range in sensitivity of the SiO
measurements could give rise to the apparent lack of correlation
between SiO luminosity and source properties.  With the exception of
Cep E, L1660 and NGC2264G, all of our sources lie within 500 pc;
furthermore 80 per cent lie within the range 330 to 450 pc. On
average, the range of noise levels ($\Delta T_{\rm mb}$) in our data
is within 26 per cent of 105 mK (although there are a couple of
sources with somewhat higher levels). Considering just the subsample
of 20 sources which lie between 330 and 450 pc the sensitivity is
uniform to within 34 per cent, assuming the dispersions in the noise
level and distance add in quadrature. The lack of correlation observed
in Figure~\ref{sioplots}a--d is therefore an intrinsic property of
these data and we conclude that our sensitivity is not a significant
contributing factor.

\subsection{SiO emission from other low-mass sources}
\label{low-mass}

A handful of other low-mass protostars are known to exhibit SiO
emission. These include IRAS 16293$-$2422 (Castets et al. 2001; Blake
et al. 1994), NGC1333$-$IRAS4 (Blake et al. 1995; Lefloch et al.
1998), BHR\,71 (Garay et al.  1998), L\,1157 (Avery \& Chiao 1996;
Bachiller et al. 2001), B1 (Yamamoto et al. 1992) and SVS13B
(Bachiller et al. 1998). While the evolutionary statuses of B1 and
SVS13B are not well determined, the others are well-known class 0
protostars. This fact lends further support to our conclusion above
that the density of the circumstellar environment plays an important
role in determining the level of SiO emission.

\subsection{Water masers and SiO emission}

Recent surveys have detected a sample of low luminosity objects which
exhibit water maser emission (Claussen et al. 1996; Furuya et
al. 2001, 2003). Four of the seven SiO sources in our sample are water
maser sources -- L\,1448-mm, NGC2024-FIR6, NGC\,2071 and HH\,212. In
addition three further objects have water masers but no SiO emission
-- RNO15FIR, HH\,7--11 and Cep E, although HH\,7--11 and Cep E have
been detected in lower-lying transitions of SiO (see section
\ref{nondet} above). In addition, of the known SiO sources that we did
not observe in this sample, three (IRAS 16293-2422, NGC1333-IRAS4 and
L\,1157) have water masers associated with them.  Furuya et al. (2001)
concluded that class 0 sources are more likely to have water masers
than more evolved sources, and suggest that this is because there is a
larger quantity of dense gas in these sources. Although the statistics
are poor, the results of our survey suggest that sources which exhibit
water maser emission may also be likely to show SiO emission.

Harju et al.\ (1998) searched for SiO emission towards water maser
sources covering a wide range of luminosities. For the range of
luminosities covered by our sample, they find a similar detection rate
($\sim$20 per cent). We also note that the detection rate for more
luminous sources increases rapidly to $\sim$50 per cent, probably
reflecting the higher temperatures and densities in these regions.

\subsection{Summary}

In practice, these factors (shock velocity, density and incomplete
beam-filling) all contribute to determining whether a source exhibits
SiO emission.  Nearby sources will suffer obviously less from
incomplete beam-filling, which may explain why we do not detect Cep E,
NGC\,2264G or L\,1660. Shock velocity and density are probably linked
as class 0 sources are found in the highest density environments, and
are observed to have the most powerful flows, while many also have
very high-velocity outflows. Therefore, it seems reasonable to
conclude that class 0 sources are those most likely to exhibit SiO
emission, and as such that SiO emission (like water masers) may be
used as a (statistical) signpost for the presence of a very young,
protostellar candidate.

\section{Conclusion}

We have observed the SiO $J$=5\too 4 line towards a sample of 25
low-luminosity (L$_* < 10^3$ L$_\odot$) protostellar outflow systems.
The line was clearly detected towards 7 of the 25 sources, a detection
rate of 28 per cent. The majority (5 out of 7) of sources detected are
of class 0 status. 
We detect a greater fraction of class 0 sources
compared with other classes (45 per cent compared with 18 and 0 per
cent for class I and II). However, given the small numbers involved,
the significance of this result should be regarded as tentative.

In four cases (HH\,25MMS, HH\,212, V-380 Ori\,NE and HH\,211) emission
was detected at positions offset from the center (in two other cases
it is known that SiO emission also occurs away from the central source
position). LTE analysis yields SiO abundances ranging from 10$^{-8}$
to as high as 8$\times$10$^{-7}$. In two cases (HH\,25MMS and HH\,212)
we have conducted LVG modeling of the emission, and find that the SiO
emission arises in regions with temperatures of typically 30--100 K,
densities of 10$^5$ to 10$^6$ cm$^{-3}$ and SiO abundances of
10$^{-11}$ to 10$^{-9}$ relative to H$_2$. These abundances are lower
than those derived from LTE analysis, a result consistent with the
notion that the SiO emission arises in clumps which are much smaller
than the 22-arcsec beam.  Higher resolution observations has shown
this is true, and we feel that further high-resolution studies are
desireable in order to determine the conditions of the post-shock gas.

We have explored a number of possible reasons for the detection of SiO
in some sources but not others, and conclude that the shock velocity
and density of the ambient medium play an important role.  We can
therefore account for the fact that most of the detected sources have
class 0 status. We conclude that while not all class 0 sources exhibit
SiO emission, SiO emission is a good signpost for class 0 sources.

\acknowledgments

The authors would like to thank PPARC for financial support and the
JCMT support staff for conducting some of the observations in service
mode. AGG was supported by NSF grant AST-9981289 to the University of
Maryland for part of this work. AGG wishes to thank Tom Hartquist for
helpful discussions on SiO chemistry in shocks, and Nicholas Chapman
for communicating results on HH\,212.  The referee is thanked for
useful comments which helped clarify the presentation. The JCMT is
operated by the Joint Astronomy Centre on behalf of the UK Particle
Physics and Astronomy Research Council, the Netherlands Organization
for Scientific Reearch and the National Research Council of Canada.

\clearpage

\begin{figure}
\figurenum{7}
\plotone{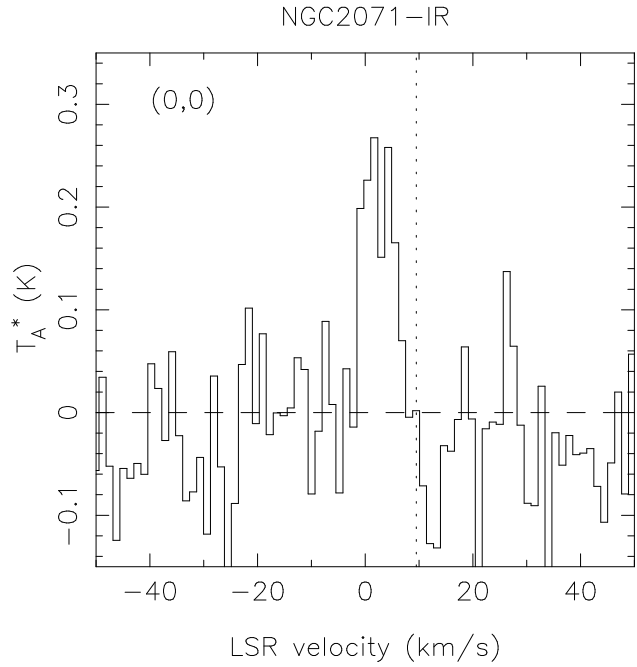}
\caption{SiO 5\too 4 spectra towards the NGC\,2071 outflow. The
  spectral resolution in this figure is 1.0 MHz. The vertical dotted
  line marks the LSR velocity.}
\label{n2071}
\end{figure}

\begin{figure}
\figurenum{8}
\plotone{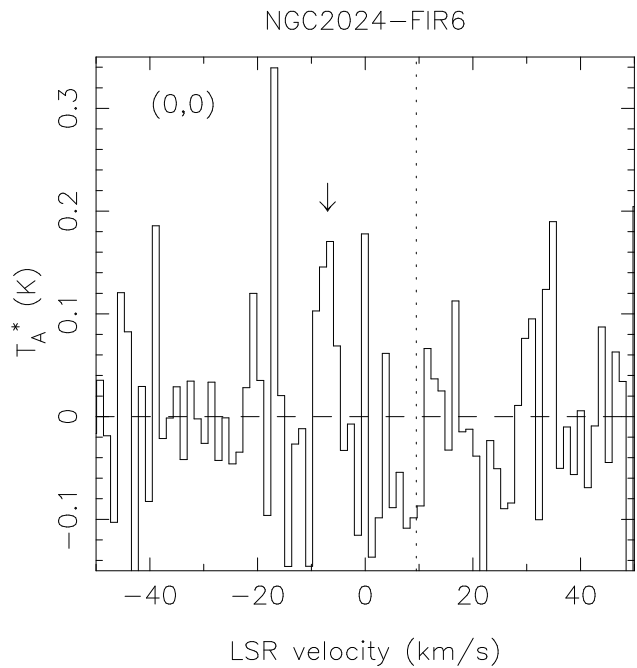}
\caption{SiO 5\too 4 spectra towards the NGC\,2024-FIR6 outflow. The
  vertical dotted line marks the LSR velocity. The tentative detection
  is marked by the small arrow at --7.2 \kms.}
\label{fir6}
\end{figure}

\clearpage

\begin{deluxetable}{lccccccc}
\tablewidth{0pt}
\tabletypesize{\small}  
\tablecaption{Source names, coordinates, distances, LSR velocities, evolutionary class and peak main-beam antenna temperatures and integrated intensities.  Where no detection was made, the 1-$\sigma$ noise level is given for a 2.5-MHz wide channel and the integrated intensity is this noise over a 12-\kms\ range (equal to the mean extent of the emission in the detected sources). \label{obs} }
\tablehead{ \colhead{Name} & \colhead{RA(1950)} &
    \colhead{Dec(1950)} & \colhead{$d$} & \colhead{$v_{\rm LSR}$} &
    \colhead{Class} &
    \colhead{Peak $T_{\rm mb}$} & \colhead{$\int T_{\rm mb}\,dv$} \\
    \colhead{} & \colhead{} & \colhead{} & \colhead{(pc)} &
    \colhead{(\kms)} & \colhead{} & \colhead{(mK)} & \colhead{(K~\kms)}}
\startdata 
Detections:\\
\objectname{L1448-mm}        & 03 22 34.3 &   +30 33 35.0 & \phn330 &   +4.0  & 0  & \phn\phm{$<$}352 & \phs17.3$\pm$1.4  \\
\objectname{HH\,211-mm}      & 03 40 48.6 &   +31 51 24.0 & \phn330 &   +9.0  & 0  & \phm{$<$}1169    & \phs22.0$\pm$1.1  \\
\objectname{V380 Ori-NE}     & 05 34 10.0 & $-$06 40 44.0 & \phn450 &   +8.5  & I  & \phn\phm{$<$}436 & \phn4.5$\pm$0.4   \\ 
\objectname{NGC\,2024-FIR6}  & 05 39 13.7 & $-$01 57 28.0 & \phn415 &   +10.0 & 0  & \phn\phm{$<$}330 & \phn0.7$\pm$0.2   \\
\objectname{HH\,212}         & 05 41 19.0 & $-$01 04 08.0 & \phn400 &   +8.0  & 0  & \phn\phm{$<$}493 & \phn3.8$\pm$0.6   \\
\objectname{HH\,25MMS}       & 05 43 33.8 & $-$00 14 36.0 & \phn400 &   +10.0 & 0  & \phn\phm{$<$}634 & \phn4.2$\pm$0.6   \\
\objectname{NGC\,2071}       & 05 44 30.7 &   +00 20 45.0 & \phn390 &   +9.0  & I  & \phn\phm{$<$}387 & \phn3.2$\pm$0.6   \\
Non-detections:\\
\objectname{B5-IRS3}         & 03 43 56.5 &   +32 33 55.0 & \phn350 &   +10.0 & II & \phn$<$106       & \phn$<$0.68 \\
\objectname{RNO\,15-FIR}     & 03 24 34.9 &   +30 02 40.0 & \phn350 &   +5.0  & 0? & \phn\phn$<$90    & \phn$<$0.58 \\
\objectname{HH\,7--11}       & 03 25 57.9 &   +31 05 50.0 & \phn350 &   +8.0  & I  & \phn\phn$<$79    & \phn$<$0.51 \\
\objectname{DG Tau}          & 04 24 01.0 &   +29 59 36.0 & \phn140 &   +6.0  & II & \phn\phn$<$87    & \phn$<$0.56 \\
\objectname{HL Tau}          & 04 28 44.4 &   +18 07 35.0 & \phn140 &   +6.5  & II & \phn$<$109       & \phn$<$0.70 \\
\objectname{L\,1634}         & 05 17 21.9 & $-$05 55 05.0 & \phn450 &   +8.0  & I? & \phn$<$159       & \phn$<$1.02 \\
\objectname{RNO\,43-mm}      & 05 29 30.4 &   +12 47 34.0 & \phn450 &   +9.5  & 0  & \phn$<$124       & \phn$<$0.79 \\ 
\objectname{HH\,34-mm}       & 05 33 03.6 & $-$06 28 49.0 & \phn450 &   +8.5  & I  & \phn$<$129       & \phn$<$0.83 \\ 
\objectname{HH\,1--2-mm}     & 05 33 57.0 & $-$06 47 55.0 & \phn450 &   +8.0  & I  & \phn$<$126       & \phn$<$0.81\\ 
\objectname{NGC\,2024-FIR5}  & 05 39 12.6 & $-$01 57 00.0 & \phn415 &   +10.0 & 0  & \phn$<$150       & \phn$<$0.96 \\
\objectname{HH\,26IR}        & 05 43 30.5 & $-$00 15 59.0 & \phn400 &   +10.0 & I  & \phn\phn$<$85    & \phn$<$0.54 \\
\objectname{SSV\,59}         & 05 43 31.1 & $-$00 15 28.0 & \phn400 &   +10.0 & I  & \phn\phn$<$86    & \phn$<$0.55 \\
\objectname{HH\,24MMS}       & 05 43 34.9 & $-$00 11 49.0 & \phn400 &   +10.0 & 0  & \phn$<$109       & \phn$<$0.70 \\
\objectname{HH110}           & 05 48 46.8 &   +02 55 01.0 & \phn400 &   +8.5  & I  & \phn\phn$<$53    & \phn$<$0.34 \\
\objectname{HH111-VLA}       & 05 49 09.3 &   +02 47 48.0 & \phn400 &   +9.0  & I  & \phn$<$126       & \phn$<$0.81 \\
\objectname{NGC\,2264G}      & 06 38 25.8 &   +09 58 52.0 & \phn800 &   +3.5  & 0  & \phn\phn$<$86    & \phn$<$0.55 \\
\objectname{L\,1660}         & 07 18 00.9 & $-$23 56 42.0 & 1500    &   +20.0 & I? & \phn$<$107       & \phn$<$0.69 \\
\objectname{Cep E}           & 23 01 10.1 &   +61 26 16.0 & \phn700 & $-$13.0 & 0  & \phn\phn$<$73    & \phn$<$0.47 \\
\enddata
\tablecomments{Units of right ascension are hours, minutes and
  seconds. Units of declination are degrees, minutes of arc and seconds of arc.}
\end{deluxetable}

\begin{deluxetable}{lccc}

\tablewidth{0pt}
\tabletypesize{\small}  

\tablecaption{Details of source coverage. Offsets are given in arcsec
  from the center position given in Table \ref{obs}. Regular grids are
  given in terms of their extent in RA and Dec respectively. Strip
  maps are quoted as their extent along the flow axis. Lateral extent
  is assumed to be 1 pixel unless stated. Positive extent is given
  along the direction of the blue lobe. The position angle (PA) for
  strip maps is given in degrees east of north. Note HH25MMS was
  covered by two different types of map. Positions in the RNO\,43-mm
  flow are taken from Bence et al.\ (1996). \label{maps} }

  \tablehead{ \colhead{Name} & \colhead{Map type} & \colhead{PA} & \colhead{Extent}
}
\startdata
\objectname{L1448-mm}        & Point & \phn\nodata &  (0,0) \\
\objectname{HH\,211-mm}      & Grid  & \phn\nodata &  $-$60 to +40, $-$20 to +30 \\
\objectname{B5-IRS3}         & Grid  & \phn\nodata &  $-$25 to +25, $-$25 to +25 \\
\objectname{RNO\,15-FIR}     & Strip & \phn+42 &  $-$100 to +100 \\
\objectname{HH\,7--11}       & Strip & +119 &  $-$100 to +100  \\
\objectname{DG Tau}          & Grid  & \phn\nodata &  $-$20 to +60, $-$20 to +0  \\
\objectname{HL Tau}          & Grid  & \phn\nodata &  $-$20 to +20, $-$20 to +0  \\
\objectname{L\,1634}         & Strip & +101 &  $-$200 to +200  \\
\objectname{RNO\,43-mm}      & Grid  & \phn\nodata &  n1 to s1, n2, n4--n6 \\
\objectname{HH\,34-mm}       & Strip & +165 &  +0 to +80 \\
\objectname{HH\,1--2-mm}     & Grid  & \phn\nodata &  $-$80 to +80, $-$80 to +80 \\
\objectname{V380 Ori-NE}     & Strip & \phn\phn+3 &  $-$80 to +80 \\
\objectname{NGC\,2024-FIR5}  & Grid  & \phn\nodata &  $-$20 to +40, $-$80 to +0 \\
\objectname{NGC\,2024-FIR6}  &       & \phn\nodata & covered in FIR5 map \\
\objectname{HH\,212}         & Strip & \phn+24 &  $-$80 to +80 \\
\objectname{HH\,26IR}        & Strip & \phn+58 &  $-$100 to +100 \\
\objectname{SSV\,59}         &       & \phn\nodata & covered in HH\,25MMS map  \\
\objectname{HH\,25MMS}       & Strip & +155 &  $-$100 to 0 \\
~                            & Grid  & \phn\nodata &  $-$20 to 80, $-$100 to 40 \\
\objectname{HH\,24MMS}       & Grid  & \phn\nodata &  $-$40 to +40, $-$20 to +0 \\
\objectname{NGC\,2071}       & Point & \phn\nodata &  (0,0) \\
\objectname{HH110}           & Strip & \phn\phn\phm{+}0 &  $-$200 to +40 \\
\objectname{HH111-VLA}       & Strip & \phn+98 &  $-$160 to +200 \\
\objectname{NGC\,2264G}      & Strip & \phn+94 &  $-$100 to +200 \\
\objectname{L\,1660}         & Strip & \phn+90 &  $-$60 to +60 \\
\objectname{Cep E}           & Grid  & \phn\nodata &  $-$20 to +20, $-$40 to +40 \\
\enddata
\end{deluxetable}

\begin{deluxetable}{lcccc}
  \tablewidth{0pt} 
  
  \tablecaption{Estimated SiO abundances towards the positions
    specified offset (in arcsec) from the center positions given in
    Table~\ref{obs}. A CO abundance of 10$^{-4}$ relative to H$_2$ has
    been assumed. CO column densities are estimated from published
    data. \label{nsio} }
  
  \tablehead{ \colhead{Source} & \colhead{Position} & \colhead{$N_{\rm
        SiO}$} & \colhead{$N_{\rm CO}$} & \colhead{$X_{\rm SiO}$}
\\
       & offset   &  (cm$^{-2}$)   & (cm$^{-2}$)  & }
\startdata
L\,1448-mm     & (0,0)       & 7.1(12) & 8.4(14) & 8.5($-$7) \\
HH\,211-mm     & ($-$10,5)   & 2.4(13) & 4.6(15) & 5.2($-$7) \\*
               & (15,$-$5)   & 1.5(13) & 3.7(15) & 4.1($-$7) \\ 
V380 Ori-NE    & (0,20)      & 5.7(12) & 2.7(16) & 2.1($-$8) \\*
               & (0,$-$20)   & 2.6(12) & 2.8(16) & 9.3($-$9) \\
HH\,212        & (0,0)       & 5.0(12) & 5.6(15) & 8.9($-$8) \\*
               & ($-$8,$-$18)& 4.3(12) & 4.0(15) & 1.1($-$7) \\
NGC\,2024-FIR6 & (0,0)       & 7.1(12) & 1.0(16) & 7.1($-$8) \\
HH\,25MMS      & (36,$-$57)  & 4.9(12) & 4.4(15) & 1.1($-$7) \\*
               & (14,$-$28)  & 4.5(12) & 1.8(16) & 2.5($-$8) \\
NGC\,2071      & (0,0)       & 4.5(12) & 2.0(16) & 2.2($-$8) \\
\enddata
\tablecomments{The notation $a(b)$ represents $a \times 10^b$.}
\end{deluxetable}

\begin{deluxetable}{lcccccccc}
\tablewidth{0pt}
\tabletypesize{\small}

\tablecaption{Summary of LVG modeling. Results are presented for each
  position (column 2) within each source. The transitions used,
  brightness temperatures and assumed velocity gradient are given in
  columns 3, 4 and 5. The parameter range for the solutions is given
  in columns 6, 7 and 8. These data are shown in
  Fig. \ref{lvgfig}a and b. The quantity $X_{\rm SiO}^{'}$
  represents the SiO abundance derived from the `Corrected' solutions
  (Fig. \ref{lvgfig}c and d). \label{lvg}}

\tablehead{
\colhead{Source} & \colhead{Position}  & \colhead{Transition} & \colhead{$T_{\rm mb}$}  &
\colhead{$dv/dr$} & \colhead{$n_{\rm H_2}$} & \colhead{$T_{\rm kin}$} & \colhead{$X_{\rm SiO}$} & \colhead{$X_{\rm SiO}^{'}$} \\
\colhead{}      & \colhead{offset}   & \colhead{} & \colhead{(K)} &
\colhead{(\kms \,pc$^{-1}$)} & \colhead{(cm$^{-3}$)} & \colhead{(K)} & \colhead{} & \colhead{}
}
\startdata
HH\,212-A      & (0,0)     &  2\too 1 & 0.23 & 350 & 0.8--2.0(5) & 50--150 & 1--3($-$9)  & 2--6($-$8)\\*
               & (0,0)     &  5\too 4 & 0.26 & \nodata & \nodata & \nodata & \nodata & \nodata \\*
HH\,212-B      & ($-$8,$-$18)&  2\too 1 & 0.60 & 175 & 0.7--1.5(5) & 50--100 & 2--5($-$9) & 5--12($-$8)\\*
               & ($-$8,$-$18)&  5\too 4 & 0.34 & \nodata & \nodata & \nodata & \nodata & \nodata \\
HH\,25MMS      & (14,$-$28) &  5\too 4 & 0.29 & 210 & 0.5--2.0(6) & 30--100 & 2--20($-$11) & 2--20($-$9)\\*
               & (14,$-$28) &  7\too 6 & 0.12 & \nodata & \nodata & \nodata & \nodata & \nodata \\*
               & (14,$-$28) &  8\too 7 & 0.07 & \nodata & \nodata & \nodata & \nodata & \nodata \\
\enddata
\tablecomments{The notation $a(b)$ represents $a \times 10^b$.}
\end{deluxetable}

\begin{deluxetable}{lcccccccc}
\tablewidth{0pt}
\tabletypesize{\scriptsize} 

\tablecaption{Details of values plotted in Fig. \ref{sioplots}. SiO
5\too 4 luminosity (derived from the brightest detection given in
Table \ref{obs}), maximum outflow velocity, momentum flux, mechanical
luminosity, source bolometric luminosity, dynamical timescale and
850-$\mu$m luminosity. If no 850 $\mu$m measurement is available,
estimates have been made from other wavelengths assuming that the flux
density is proportional to $\nu ^3$ (Chandler \& Richer 2000).
\label{plotdata}}

\tablehead{
Source & $L_{\rm SiO 5\rightarrow 4}$   & $v_{\rm max}$ & $dp/dt$ & $L_{\rm mech}$ &
$L_{\rm bol}$ & $t_{\rm dyn}$ & $L_{\rm 850\mu m}$ & References\\ 
 &  (L$_\odot$) & (km\,s$^{-1}$) & (M$_\odot$
 km\,s$^{-1}$\,yr$^{-1}$) & (L$_\odot$) & (L$_\odot$) & (yr) & (L$_\odot$\,GHz$^{-1}$) & 
}
\startdata
L1448-mm      &  \phm{$<$}5.4($-$9)  &  70      & 2.3($-$4)  &  0.55      &    8    & 3.5(3)   & 1.3($-$4)   & 1,2,3,4,5 \\
HH211-mm      &  \phm{$<$}6.8($-$9)  &  40      & 6.7($-$5)  &  1.3($-$3) &   15    &  750     & 2.1($-$4)   & 3,6 \\
B5-IRS3       &  $<$2.3($-$10)       &   5      & \nodata    & \nodata    & \nodata & \nodata  & \nodata    & 5 \\ 
RNO15-FIR     &  $<$2.0($-$10)       &   7      & 4.0($-$6)  &  2.5($-$3) &   12    & 3.1(4)   & \nodata    & 5,7,8 \\ 
HH7-11        &  $<$1.7($-$10)       &  27      & 2.2($-$4)  &  0.47      &   50    & 5.6(3)   & 2.4($-$4)   & 3,5,9,10 \\ 
DG-Tau        &  $<$3.1($-$11)       &   3.5    & 1.9($-$6)  &  5.4($-$4) &   1     & 3.4(4)   & 6.1($-$7)   & 5,11 \\ 
HL-Tau        &  $<$3.9($-$11)       &   7      & 1.7($-$5)  &  9.0($-$3) &   7     & 9.0(3)   & 1.8($-$5)   & 3,5,12,13 \\ 
L1634         &  $<$5.9($-$10)       &   5      & \nodata    & \nodata    &  17     & \nodata  & \nodata    & 14 \\ 
RNO43-mm      &  $<$4.5($-$10)       &   5      & 2.7($-$5)  &  0.03      &   6     & 3.3(4)   & \nodata    & 5,15 \\ 
HH34-mm       &  $<$4.8($-$10)       &   5      & 3.7($-$5)  &  0.01      &  35     & 1.3(4)   & 7.9($-$5)   & 16,17,18 \\ 
HH1-2mm       &  $<$4.5($-$10)       &   3      & 8.1($-$4)  &  0.2       &  61     & 1.1(4)   & 3.6($-$5)  & 5,16,19 \\ 
V380-Ori-NE   &  \phm{$<$}2.6($-$9)  &  13      & 1.6($-$4)  &  0.17      &  10     & 1.6(4)   & 7.9($-$5)   & 20 \\ 
NGC2024-FIR5  &  $<$4.8($-$10)       &  20      & 1.8($-$4)  &  3         & \nodata & 5.0(4)   & 5.8($-$4)   & 21,22,23 \\ 
NGC2024-FIR6  &  \phm{$<$}3.4($-$10) &  50      & 5.2($-$3)  &  15        &  3(3)   & 400      & 3.6($-$4)   & 21,23,24 \\ 
HH212         &  \phm{$<$}1.7($-$9)  &  20      & \nodata    & \nodata    &  14     & 1.4(4)   & 2.0($-$5)   & 25,26 \\
HH26IR        &  $<$2.5($-$10)       &  18      & 9.2($-$6)  &  7.1($-$3) &   5     & 1.3(4)   & 2.9($-$5)   & 27,28 \\ 
SSV59         &  $<$2.5($-$10)       &   5      & \nodata    & \nodata    &   5     & \nodata  & 1.9($-$5)   & 27,28 \\
HH25MMS       &  \phm{$<$}1.9($-$9)  &   8      & 2.8($-$5)  &  0.01      &   6     & 1.7(4)   & 6.7($-$5)   & 28,29 \\ 
HH24MMS       &  $<$3.1($-$10)       &   7.5    & 5.1($-$4)  &  0.3       &  10     & 3.7(4)   & 1.7($-$4)   & 5,28,30 \\ 
NGC2071       &  \phm{$<$}1.4($-$9)  &  38      & 2.9($-$4)  &  1.8       & 600     & 2.0(4)   & 1.4($-$3)   & 5,31,32 \\ 
HH110         &  $<$1.5($-$10)       &  10      & 5.7($-$4)  &  1.2       &   7     & 5.0(4)   & \nodata    & 5,33 \\ 
HH111-VLA     &  $<$3.7($-$10)       &  10      & \nodata    & \nodata    &  25     & \nodata  & 7.3($-$5)   & 5,34,35 \\
NGC2264G      &  $<$9.9($-$10)       &  60      & 3.2($-$4)  &  1.0       &  12     & 4.0(4)   & 1.2($-$4)   & 5,36,37,38 \\ 
L1660         &  $<$4.2($-$9)        &   4.5    & 6.6($-$4)  &  0.24      & 100     & 2.2(4)   & \nodata    & 5,39 \\ 
Cep-E         &  $<$6.5($-$10)       &  20      & 3.4($-$5)  &  0.06      &  85     & 2.5(4)   & 9.4($-$5)  & 40 \\ 
\enddata
\tablecomments{The notation $a(b)$ represents $a\times 10^b$.}
\tablerefs{
(1) Bachiller et al.\ 1990; 
(2) Bachiller et al.\ 1995; 
(3) Chandler \& Richer 2000; 
(4) Shirley et al.\ 2000; 
(5) Wu, Huang \& He 1996; 
(6) Gueth \& Guilloteau 1999; 
(7) Davis et al.\ 1997b; 
(8) Meehan et al.\ 1998 
(9) Knee \& Sandell 2000; 
(10) Bachiller \& Cernicharo 1990; 
(11) Mitchell et al.\ 1994; 
(12) Looney et al.\ 2000; 
(13) Monin, Pudritz \& Lazareff 1996; 
(14) Davis et al.\ 1997a; 
(15) Bence et al.\ 1996; 
(16) Chernin \& Masson 1995; 
(17) Stapelfeldt \& Scoville 1993; 
(18) Cohen \& Schwartz 1987; 
(19) Chini et al.\ 2001; 
(20) Davis et al.\ 2000; 
(21) Chernin 1996; 
(22) Richer et al.\ 1989; 
(23) Visser et al.\ 1998; 
(24) Richer 1990; 
(25) Lee et al.\ 2000; 
(26) Zinnecker et al.\ 1992; 
(27) Gibb \& Heaton 1993; 
(28) Phillips, Gibb \& Little 2001; 
(29) Gibb \& Davis 1998; 
(30) Chini et al.\ 1993; 
(31) Moriarty-Schieven et al.\ 1989; 
(32) Mitchell et al.\ 2001; 
(33) Reipurth \& Olberg 1991; 
(34) Hatchell et al.\ 1999; 
(35) Cernicharo \& Reipurth 1996; 
(36) Lada \& Fich 1996; 
(37) Fich \& Lada 1997; 
(38) Ward-Thompson et al.\ 1995;  
(39) Schwartz, Gee \& Huang 1988; 
(40) Ladd \& Hodapp 1997. 
}
\end{deluxetable}



\end{document}